\documentclass[10pt]{article}
\usepackage{amsmath}
\usepackage{amssymb}
\usepackage{graphicx}
\usepackage{cite}

\let\tsection\section
\renewcommand{\section}{\setcounter{equation}{0}\tsection}

\def\F{{\cal F}}

\def\rhobar{\bar{\rho}}
\newcommand{\KWASEP}{K_{\lambda}(\rho_a,\rho_b)}
\newcommand{\ymax}{y_{\text{m}}}
\newcommand{\rvec}[1]{| #1 \rangle }
\newcommand{\lvec}[1]{\langle #1 |}
\newcommand{\Gm}[1]{{\cal G}\big( #1\big)}

\newcommand{\Gbis}[1]{{\cal H}\left( \vphantom{\big(} #1\right)}
\newcommand{\D}{\mathbf{D}}
\newcommand{\E}{\mathbf{E}}

\newcommand{\V}{|V\rangle}
\newcommand{\n}{|n\rangle}
 
\newcommand{\W}{\langle W|}

\begin{document}
%\today
\begin{center} 

 LARGE DEVIATION FUNCTIONAL OF 
THE WEAKLY  ASYMMETRIC EXCLUSION PROCESS

\vskip10pt
C. Enaud 
and B. Derrida
\vskip .5cm
Laboratoire de Physique Statistique \footnote{email:  enaud@lps.ens.fr
and derrida@lps.ens.fr}, \\
Ecole Normale Sup\'erieure, 24 rue Lhomond, 75005 Paris, France 

\today
\vskip20pt

\end{center} 

\vskip20pt
\noindent {\bf Abstract}
We obtain the large deviation functional of a density profile for the
asymmetric exclusion process of $L$ sites with open boundary conditions
when the asymmetry scales like $\frac{1}{L}$. We recover as limiting
cases the expressions derived recently for the symmetric (SSEP) and the
asymmetric (ASEP) cases. In the ASEP limit, the non linear differential
equation one needs to solve can be analysed by a method which resembles
the WKB method.

\vskip10pt
\noindent
{\bf Key words:} Large deviations, asymmetric simple exclusion process, open
system, stationary non-equilibrium state.
\newpage
%\tableofcontents
\section{Introduction}
The study of steady states of non-equilibrium systems has motivated
a lot of works over the last decades \cite{dGM,Ligg1, HS, HS1, EKKM,Mal, SCHMITT,
SCHMITT2, KRUG}.
It is now well established that
non-equilibrium systems exhibit in general long-range correlations
in their steady state \cite{HS1, ZS,DE, DEL}.

One of the most studied examples of non-equilibrium system is the one
dimensional exclusion process with open boundaries \cite{HS1, ZS, Ligg2}.
The system is a one dimensional lattice gas on a lattice of $L$ sites.
At any given time, each site  ($1 \leq i \leq L$) is either empty or
occupied by at most one particle and the system evolves according to
the following rule:
 in the interior of the system ($2
\leq
i \leq L-1$), during each infinitesimal time interval $dt$, a particle attempts to jump to its right neighboring site
with
probability $dt$  and to its left neighboring site with probability $q \  dt$. 
The jump is completed if the target site is empty, otherwise nothing
happens. The parameter $q$ represents a bias (i.e. the effect of an external field
in the bulk).
The boundary sites $i=1$ and $i=L$ are connected to  reservoirs of particles
and
their dynamics is modified as follows: if site 1 is empty, it  becomes
occupied with probability  $\alpha \  dt$ by a particle from the left
reservoir, and  if it is
occupied, the particle is removed with a probability $\gamma \   dt$ or attempts to jump to site 2 (succeeding if this
site is
empty) with probability $dt$.  Similarly, if site $L$ is occupied, the particle may
either jump out of the system (into the right reservoir) with probability
$\beta \  dt$
or
to site $L-1$  with probability  $q \ dt$, and if it is empty, it becomes
occupied with probability $\delta \  dt$.

The rates $\alpha$, $\beta$, $\gamma$ and $\delta$  at which particles are injected at sites $1$ and $L$
 can be thought as the contact of the chain with a
reservoir of particles at density $\rho_a$  at site $1$ and  with a
reservoir at density $\rho_b$ at site $L$.
The reservoir densities  $\rho_a$ and $\rho_b$ are related to $\alpha$,
$\beta$, $\gamma$ and $\delta$  by 
(see appendix)
\begin{subequations}
\begin{align}
\rho_a&=\frac{1-q+\alpha+\gamma-\sqrt{(\alpha-\gamma-1+q)^2+4\alpha
\gamma}}{2(1-q)} \label{rhoa}\\
\rho_b&=\frac{1-q-\beta-\delta+\sqrt{(\beta-\delta-1+q)^2 +4\beta
\delta}}{2(1-q) \label{rhob}}
\end{align}
\end{subequations}

For $q=1$, the bulk dynamics is symmetric and the model is called the
{\it Symmetric Simple Exclusion
Process} (SSEP) \cite{ Ligg1,HS}. In the steady state, there is a current of particles
flowing from one reservoir to the other  (when $\rho_a\neq \rho_b$) and the steady state density profile
is linear.

For
 $0 \leq q < 1$, the bulk dynamics is asymmetric and the model is called
the
{\it Asymmetric Simple Exclusion
Process} (ASEP) \cite{DEHP, Sas, BECE}. When the densities $\rho_a$ and $\rho_b$ 
  vary, the system
exhibits phase transitions, with different phases: a low density phase, a
high density phase and a maximal current phase \cite{BECE,Sandow, PS, SD}. On a
macroscopic scale,
the steady state profile is constant except along the first order
transition line $ \rho_a=\rho_b< 1/2$ 
 separating the low and the high density phases.

In the large $L$ limit, the probability $P_L\big(\{\rho(x)\}\big)$ of
observing  a given
macroscopic density profile $\rho(x)$, $0 \leq x \leq 1$ can be expressed
through the large deviation functional $\F (\{\rho(x)\};\rho_a, \rho_b)$
by 
\begin{align} \label{PLFint}
P_L\big(\{\rho(x)\}\big) &\sim e^{-L \F (\{\rho(x)\};\rho_a, \rho_b)}
\end{align}
This large deviation functional  $\F$ is an extension of the notion of free
energy to non-equilibrium systems \cite{DLS, BDGJL, KOV}. \\
One can think of a number of distinct definitions of
$P_L\big(\{\rho(x)\}\big)$ which all lead to the same $\F$ in the large
$L$ limit. Here, by 
 dividing the
system into $k$ boxes of size $L_1$, $L_2$,\dots, $L_k$ (with $\sum_{i=1}^k
L_i =L$), we define $q_{L_1,
L_2,\dots, L_k}(N_1,N_2,\dots, N_k)$ as the probability of observing in
the steady state $N_1$ particles in the first box, $N_2$ in the second,
\dots, $N_k$ in the last box. Then if we identify  $P_L(\{\rho(x)\})$
with $q_{L_1,
L_2,\dots, L_k}(N_1,N_2,\dots, N_k)$, one expects that for large $L$,
(\ref{PLFint}) holds  when $1 \ll L_i \ll L$ and $N_i$ is
the integer part of $L_i \rho(x_i)$ where $x_i=\sum_{j=1}^i \frac{L_j}{L}$

In \cite{DLS,DLS2,BDGJL,BDGJL2,DLSasep,DLSasep2}, the following exact expressions of ${\cal F}
(\{\rho(x)\};\rho_a,\rho_b)$ were obtained:

 In the symmetric case, i.e. for
$q=1$, it was shown in \cite{DLS,BDGJL} that 
\begin{eqnarray}
 {\cal F}_{\text{SSEP}}(\{\rho(x),\rho_a,\rho_b\})
 &=& \sup_{F(x)} \int_0^1 dx \left\{ \rho(x) \log \left(\frac{ \rho(x)}{
 F(x)}\right)
   + (1- \rho(x)) \log \left(\frac{ 1 - \rho(x)}{ 1 -F(x)}
   \right)\right.
     \nonumber\\
 && \hskip100pt
  + \left.\log \left(\frac{ F'(x) }{ \rho_b - \rho_a} \right) \right\}.
\label{simple_expression}
\end{eqnarray}
where the $\sup$ is over all monotone functions $F(x)$ which satisfy
 \begin{equation}
 F(0)= \rho_a \;,  \ \ \ \ \; \ \ \ \  F(1)= \rho_b \;.
\label{condition5}
 \end{equation}
The auxiliary function $F(x)$ which achieves the $\sup$ is the monotone
solution of the nonlinear differential equation 
 \begin{equation}
\rho(x) =
 F(x) + \frac{F(x)(1-F(x)) F''(x)}{F'(x)^2}\; ,
\label{rho(t)}
 \end{equation}
with the boundary conditions (\ref{condition5}).

In the asymmetric case (i.e. for $q<1$), the expression of
${\cal F}
(\{\rho(x)\};\rho_a,\rho_b)$ is given  \cite{DLSasep,DLSasep2} 
\begin{itemize}
\item{in the case $\rho_a\ge\rho_b$ by}
 \begin{eqnarray}
\label{result}
 {\cal F}_{\rm ASEP} (\{\rho(x)\};\rho_a,\rho_b)
   = -  K(\rho_a,\rho_b)  \hskip-190pt &&  \\
\nonumber &+& \sup_{F(x)} \int_0^1 dx\,
\rho(x) \log \left[ \rho(x) (1- F(x)) \right] +
(1 - \rho(x)) \log \left[ (1- \rho(x))  F(x) \right],
 \end{eqnarray}
 where the $\sup$ is over all {\it monotone non-increasing} functions
$F(x)$ such that $F(0)=\rho_a$ and $F(1)=\rho_b$ and
 \begin{equation}
 K(\rho_a,\rho_b) =
 \sup_{ \rho_b \leq \rho \leq \rho_a}  \log [ \rho (1 - \rho) ],
\label{Kdef1}
 \end{equation}
As shown in \cite{DLSasep2}, the function $F(x)$ which gives the sup in
(\ref{result}) is the derivative of the concave envelope of
$\int_0^x[1-\rho(x')]dx'$, whenever this derivative belongs to $]\rho_b,
\rho_a[$, and it takes the value $\rho_a$ or $\rho_b$ otherwise. As a
result, when $F(x)$ differs from $\rho_a$ and $\rho_b$, it is made up of
a succession of domains where $F(x)=1-\rho(x)$ and of domains
where $F(x)$ is constant (as $F(x)$ is decreasing, it cannot in general
coincide with $1-\rho(x)$ everywhere). In the domains where  $F(x)$ is
constant (and differs
from $\rho_a$ and $\rho_b$) it satisfies a
Maxwell construction rule: when $F(x)=C$  for $0<t\leq
x\leq u<1$, its  value  is determined by
\begin{align} \label{maxwell}
(u-t) C &= \int_t^u \big[1-\rho(x)\big] \, dx
\end{align}
\item and in the case $\rho_a\le\rho_b$ by \\ 
  \begin{eqnarray}
{\cal F}_{\rm ASEP} (\{ \rho(x) \};\rho_a,\rho_b)) =
-  K(\rho_a,\rho_b) +
\hskip100pt
\label{result3} \\
   \inf_{0 \leq y \leq 1}
 \left\{
 \int_a^y dx\, \rho(x) \log \left[{ \rho(x)   (1 -\rho_a)} \right]
+ (1 - \rho(x)) \log \left[( 1- \rho(x))    \rho_a \right]
\right.
 \nonumber \\
\left.
  +   \int_y^b dx\, \rho(x) \log \left[ \rho(x)  (1 -  \rho_b) \right]
+ (1 - \rho(x)) \log \left[ (1- \rho(x))    \rho_b \right]
 \right\} .
\nonumber
 \end{eqnarray}
where
 \begin{equation}
K(\rho_a,\rho_b) = \min \left[  \log  \rho_a (1 - \rho_a)
, \log \rho_b(1-\rho_b)  \right],
\label{Kdef2}
 \end{equation}
\end{itemize}

The goal of the present paper is to reconcile the expression valid in
the symmetric case (\ref{simple_expression})
with those valid in the asymmetric case (\ref{result},\ref{result3}) by calculating
the large deviation functional
in a weak asymmetry regime, which interpolates between the two, where $q
\to 1$ as $L \to \infty$
with
$q= 1 - \frac{\lambda}{L}$. The SSEP and the ASEP appear therefore as
limiting cases of the results obtained in the present paper.

The outline of the paper is as follows:
in section 2, we summarize our results by writing several equivalent
expressions of $\F$ in the weak asymmetry regime. %for this functional.
In section 3 we give the details of our derivation.
In section 4, we 
show how the SSEP and the ASEP expressions can be recovered in the limit
$\lambda \to 0$ and 
$\lambda \to \infty$. 
 The large $\lambda$ limit  is
somewhat reminiscent of the WKB method and the calculation of the
position (\ref{maxwell})
of the plateaux in the Maxwell construction of the function $F(x)$  have an origin very similar
to the Bohr-Sommerfeld rule in the WKB method \cite{WKB, MathMet}.
In section 5 we extend the range of validity of our results and show in
particular that they remain true when detailed balance is verified.

\section{Main results}

We consider here a  weak asymmetry regime defined as a situation  where $q \to 1$ as $L \to
\infty$,
\begin{align} \label{WASEPdef}
q&=1-\frac{\lambda}{L}
\end{align}
keeping $\lambda$ fixed.

For technical reasons which will become clear at the end of section 3.3,
our results are limited to the case
\begin{align} \label{val_dom1}
\lambda&>0 &&\text{and}& \rho_a&>\rho_b
\end{align}
In section 5, we will discuss some extensions to a broader range of
parameters.

Our main result is that in the weak asymmetry regime, the large deviation
 functional is given by
\begin{multline} \label{FWASEPwithy}
\F(\{\rho(x)\};\rho_a, \rho_b)= -\KWASEP +\inf \limits  _{\{y(x)\}}
\left\{ 
     \vphantom{-y(0)\log\frac{1-\rho_a}{\rho_a}-y(1) \log
        \frac{\rho_b}{1-\rho_b}+
        \int_{0}^{1}dx \left[-\log \frac{\left|1-e^{-\lambda y}\right|}{\lambda}+\rho \log \rho
        +(1-\rho)\log{(1-\rho)}   +(1-\rho+y')
        \log{(1-\rho+y')}+(\rho-y')\log{(\rho-y')} \right]}
   y(0)\log\frac{\rho_a}{1-\rho_a}+y(1) \log \frac{1-\rho_b}{\rho_b} \right. \\ 
  +\int_{0}^{1}dx \left[-\log \frac{1-e^{-\lambda y}}{\lambda}+\rho \log \rho
  +(1-\rho)\log{(1-\rho)} 
  \vphantom{-\log \frac{1-e^{-\lambda y}}{\lambda}+\rho \log \rho
      +(1-\rho)\log{(1-\rho)}+(1-\rho+y')
      \log{(1-\rho+y')}+(\rho-y')\log{(\rho-y')}} 
  \right. \\ 
  \left. \left.
   \vphantom{-\log \frac{1-e^{-\lambda y}}{\lambda}+\rho \log \rho
    +(1-\rho)\log{(1-\rho)}+(1-\rho+y')
    \log{(1-\rho+y')}+(\rho-y')\log{(\rho-y')}}+(1-\rho+y')
    \log{(1-\rho+y')}+(\rho-y')\log{(\rho-y')} 
   \right]
   \vphantom{-y(0)\log\frac{1-\rho_a}{\rho_a}-y(1) \log
     \frac{\rho_b}{1-\rho_b}+
     \int_{0}^{1}dx \left[-\log \frac{1-e^{-\lambda y}}{\lambda}+\rho \log \rho
     +(1-\rho)\log{(1-\rho)}   +(1-\rho+y')
     \log{(1-\rho+y')}+(\rho-y')\log{(\rho-y')} \right]} 
    \right\}
\end{multline}
where the $\inf$ is  over all continuous positive functions
$y(x)$ satisfying 
\begin{align}
\rho(x)-1 \leq y'(x) \leq \rho(x)\ \ . \notag
\end{align} 
We will show at the end of section 3.5 that the constant $\KWASEP$ is given by
\begin{align} \label{KWASEPJ}
\KWASEP&=\log(J)-\int_{\rho_b}^{\rho_a}\frac{d\rho}{\lambda
\rho(1-\rho)}\log\left(1-\frac{\lambda \rho(1-\rho)}{J}\right)
\end{align}
 where the parameter  $J$ is  solution of 
\begin{align} \label{current}
\int_{\rho_b}^{\rho_a}\frac{d\rho}{J-\lambda \rho(1-\rho)}&=1 \ \ \ .
\end{align}
The parameter $J$ is in fact related to the steady state current $j$ by
(see section 3.5) 
\begin{align} \label{Jdef}
J&=\lim_{L \rightarrow \infty} L j
\end{align}

Expression (\ref{FWASEPwithy}) for $\F$ can be rewritten  in a form  which interpolates between the symmetric 
(\ref{simple_expression}) and the asymmetric (\ref{result}) cases (see
section
\ref{der_FWASEPwithf}):

\begin{multline} \label{FWASEPwithf}
\F(\{\rho(x)\};\rho_a, \rho_b)=-\KWASEP+\int_0^1dx\left\{ 
      \vphantom{ \rho\log\frac{\rho}{F}+(1-\rho)\log\frac{1-\rho}{1-F}
+\log(|F(1-F)\lambda-F'|)+\frac{F'}{\lambda F(1-F)}\log \left(
-\frac{F'}{F(1-F)\lambda-F'}\right)} 
      \rho\log{\frac{\rho}{F}}+(1-\rho)
      \log\frac{1-\rho}{1-F} \right.\\
 \left. \vphantom{\rho\log\frac{\rho}{F}+(1-\rho)
            \log\frac{1-\rho}{1-F}
           +\log(F(1-F)\lambda-F')+\frac{F'}{\lambda
           F(1-F)}\log \left( -\frac{F'}{F(1-F)\lambda-F' }\right) }
+\log{(F(1-F)\lambda-F')}+\frac{F'}{\lambda
F(1-F)}\log{\left(-\frac{F'}{F(1-F)\lambda-F'} \right)}\right\}
\end{multline}
where the function $F(x)$ is  the solution of the differential
equation
\begin{align} \label{f_WASEP}
 (F-\rho){F'}^2+ F(1-F)F'' +\lambda
F(1-F)(F-1+\rho)F'=0
\end{align}
with the boundary conditions
\begin{align} \label{f_WASEPBC}
F(0) &= \rho_a&
F(1) &= \rho_b
\end{align}
In the range of validity of our derivation (\ref{val_dom1})  this differential equation has a unique solution, and
this solution is monotone (see section \ref{der_FWASEPwithf}).\\
Actually, if we consider the right hand side of (\ref{FWASEPwithf}) as a
functional of function $F$, then  (\ref{f_WASEP})
appears to be the condition that $F$ maximizes this functional under the
constraint (\ref{f_WASEPBC}) (see the end of section \ref{der_FWASEPwithf}), leading to
\begin{multline} \label{Ffmax}
\F(\{\rho(x)\},\rho_a, \rho_b)=\sup_F \left[-\KWASEP+\int_0^1dx\left\{
      \vphantom{ \rho\log\frac{\rho}{F}+(1-\rho)\log\frac{1-\rho}{1-F}
+\log(|F(1-F)\lambda-F'|)+\frac{F'}{\lambda F(1-F)}\log \left(
-\frac{F'}{F(1-F)\lambda-F'}\right)}
      \rho\log{\frac{\rho}{F}}+(1-\rho)
      \log\frac{1-\rho}{1-F} \right. \right. \\
 \left. \left. \vphantom{\rho\log\frac{\rho}{F}+(1-\rho)
            \log\frac{1-\rho}{1-F}
           +\log(F(1-F)\lambda-F')+\frac{F'}{\lambda
           F(1-F)}\log \left( -\frac{F'}{F(1-F)\lambda-F' }\right) }
+\log{(F(1-F)\lambda-F')}+\frac{F'}{\lambda
F(1-F)}\log{\left(-\frac{F'}{F(1-F)\lambda-F'} \right)}\right\} \right]
\end{multline}
where the $\sup$ is over  all decreasing functions $F$ satisfying
(\ref{f_WASEPBC}).

As $\F$ is a $\sup$ over convex functions of $\rho$, it is a convex
function of $\rho$ in domain (\ref{val_dom1}).

A by-product of  (\ref{FWASEPwithf}) 
 is  (see section \ref{mlp}) that  the most likely
profile $\rhobar(x)$ is solution of
\begin{align} \label{solrhobar}
\int_{\rhobar(x)}^{\rho_a}\frac{d\rho}{J-\lambda \rho(1-\rho)}&=x \ \ \ .
\end{align}
Depending on the boundary conditions this leads either to a $\tan$ profile,
 a $\tanh$ profile, or a $\coth$ profile.

\section{Derivation \label{der}}
\subsection{The matrix method}

The equal time steady state properties of the ASEP can be exactly calculated using
the so-called matrix method \cite{DEHP}. 
Let us consider a  microscopic configuration defined by its occupation
number $\{\tau_i\}$ where $\tau_i=1$ when site $i$ is occupied by a
particle, and $0$ otherwise. It can be shown   that  the
steady state probability of such a configuration for a lattice of $L$
sites can be written as
\begin{align} \label{Ptaui}
P(\{\tau_i\})&=\frac{\W \prod_{i=1}^{L} (\D \tau_i +\E
(1-\tau_i))\V}{Z_L(q)}
\end{align}
with $Z_L(q)$ being a normalization factor defined by
\begin{align} \label{ZL}
Z_L(q)&=\W (\D+\E)^L \V
\end{align}
where $\D$ and $\E$ are two operators fulfilling the followings
algebraic rules:
\begin{subequations}
\begin{align}
\D \E - q \E \D &=  \D + \E  \;,\label{alg1}\\
  \{\beta \D-\delta \E\} \V &=   \V\;,\label{alg2}\\
   \W\{\alpha \E-\gamma \D\}  &=   \W\;\label{alg3}.
\end{align}
\end{subequations}
These rules (\ref{alg1})-(\ref{alg3}) allow the computation of all
equal time steady state properties without the need of finding an
explicit representation. 

The two point correlation function  $<\tau_i
\tau_j>$ (where the symbol $<.>$ stands for the average  with respect to
the steady state probability) is given by
\begin{align} \label{corr_func}
<\tau_i \tau_j> &= \frac{\W (D+E)^{i-1} D (D+E)^{j-i-1} D
(D+E)^{L-j}\V}{Z_L(q)}
\end{align}
and the steady state
current $j$ between site $i$ and  $i+1$
is given by:
\begin{align} \label{j_corr_finc}
j&=<\tau_i (1-\tau_{i+1})>-q<(1-\tau_i)\tau_{i+1}>
\end{align}

Using expression (\ref{corr_func}) for the correlation function and the
algebra rule (\ref{alg1}), one gets
\begin{align} \label{current_Z}
j&=\frac{Z_{L-1}(q)}{Z_L(q)} \ \ \ .
\end{align}
Clearly the current does not depend on the site $i$, as it should, due to
the conservation of the number of particles.

If we divide the system of size $L$ in $k$ boxes of size
$L_1$, $L_2$, \dots $L_k$ the probability  of finding $N_1$ particles in
the first box, $N_2$ in the second, \dots and $N_k$ in the last box is given by%, with
\begin{align} \label{PLmatrix}
q_{L_1, L_2, \cdots L_k}(N_1,N_2, \cdots, N_k)&=\frac{\W X_{ L_1}(N_1)
X_{L_2} (N_2) \cdots X_{L_k}(N_k)
\V}{Z_L(q)}
\end{align}
where $X_l(N)$ is the sum over all products of $l$ matrices containing
exactly $N$ matrices $\D$ and $l-N$ matrices $\E$.

\subsection{A representation for $\D$ and $\E$}

All physical quantities such as (\ref{corr_func}), (\ref{j_corr_finc}) or
(\ref{PLmatrix}) do not depend on the representation of the matrices $\D$
and $\E$ and of the vectors $\V$ and $\W$ which satisfy the algebra
(\ref{alg1})-(\ref{alg3}). Several representations have been used to
solve (\ref{alg1}-\ref{alg3}) \cite{Sandow, Sas, BECE, ER, DEHP}.
We choose  in this section a particular representation which will be
convenient for the remaining of our derivation.

 If we write the operators $\D$ and $\E$ as infinite matrices of the
form
\begin{subequations}
\begin{align} \label{repD}
 \D &= \frac{1}{1-q}\left[ \begin{array}{c c  c c c}
        1-d &  1-q  & 0      &       &\cdots      \\
          0 & 1-d q & 1-q^2  &  0    &\cdots       \\
           0&      0& 1-dq^2 & 1-q^3 &0       \\
             &\ddots & \ddots & \ddots& \ddots
             \end{array} \right], \\
 \E &= \frac{1}{1-q}\left[ \begin{array}{c c c c c }
      1-e & 0 & 0 & 0 & \cdots \\ 1-ed & 1-e q & 0 & 0&\cdots \\
     0 & 1-edq & 1- eq^2&0&\cdots\\
     0 &\ddots&\ddots&\ddots&\ddots
             \end{array} \right], \label{repE}
\end{align}
\end{subequations}
we find that they satisfy the algebraic rule (\ref{alg1}) for arbitrary
choices of $d$ and $e$.

Let call $\{\n\}_{n\ge 1}$ the vector of the associated basis.
If we look for vectors $\V$ and $\W$ of the form 
\begin{subequations}
\begin{align}
\W &= \sum_{n=1}^{\infty}
\left(\frac{1-\rho_a}{\rho_a}\right)^n\langle n| \label{W}\\
\V &=\sum_{n=1}^{\infty}
\left(\frac{\rho_b}{1-\rho_b}\right)^n\frac{(ed;q)_{n-1}}{(q;q)_{n-1}}
|n \rangle \label{V}
\end{align}
\end{subequations}
where  $\rho_a$ and $\rho_b$ are for the moment arbitrary and the symbol
$(x;q)_i$ stands for the $q$-shifted factorial
defined by $(x;q)_0=1$ and
\begin{subequations}
\begin{align}
(x;q)_i&=\prod_{k=0}^{i-1}(1-x q^k)&
\text{for }k&>0 \ \ \ 
\end{align}
\end{subequations}
one can check that (\ref{W}) and (\ref{V}) fulfill the algebraic rules (\ref{alg2}) and (\ref{alg3}) if
the parameters $\rho_a$, $\rho_b$, $d$ and $e$ satisfy:
\begin{subequations}
\begin{align}
\frac{\alpha}{\rho_a}-\frac{\gamma}{1-\rho_a}&=1-q
\label{eq_rhoa}\\
\frac{\beta}{1-\rho_b}-\frac{\delta}{\rho_b}&=1-q
\label{eq_rhob}
\end{align}
\begin{align} \label{d}
d&=\frac{\delta (1-\rho_b)}{\beta \rho_b}
\intertext{and}
e&=\frac{\gamma \rho_a}{\alpha (1-\rho_a)} \label{e}
\end{align}
\end{subequations}
Note that the parameters $\rho_a$ and $\rho_b$ defined by 
(\ref{rhoa}) and (\ref{rhob}), that we interpreted as the reservoir
densities, are the unique  solutions of the equations  (\ref{eq_rhoa})
and (\ref{eq_rhob})  such that $0 \leq \rho_a \leq 1$ and $0 \leq
\rho_b \leq 1$ (this is why we use the same symbol in (3.9) and
(1.1)).

If we use this representation when $q<1$ (which is not a restriction due
to the left-right symmetry), we see that the condition for $\W X \V$ to
be finite  when $X$ is an arbitrary product of $\D$ and $\E$ is that
$\rho_a>\rho_b$, leading to (\ref{val_dom1}).
In section \ref{extension}, we will show that this representation remains
valid for some part of the domain $q>1$, thus allowing us to extend the
result of section 2 .

\subsection{The sum over paths and the derivation of (\ref{FWASEPwithy}) \label{der_FWASEPwithy}}

The basic idea of our derivation of (\ref{FWASEPwithy}) is to expand the matrix products such as
(\ref{Ptaui}) or (\ref{PLmatrix}) as a sum over paths \cite{DEL}, in much the
same way as the path integral formulation of  quantum mechanics.

Consider the set of discrete walks $w$ of
$k$ steps. 
Let $n_i(w)$ ($0 \leq i\leq k$) be the integer position of the walk after the $i$th step.
The walks we consider remain positive $n_i>0$ and their increment at each
step satisfies
\begin{align} \label{con_disc_walk}
n_i&> 0&&\text{and}&-L_i \leq n_i-n_{i-1} \leq L_i
\end{align}
From (\ref{PLmatrix}), we  deduce
\begin{align} \label{discretwalk}
q_{L_1,L_2,\cdots L_k}\big(N_1,N_2, \cdots,N_k\big)&= \frac{1}{Z_L(q)}\sum_{w} \W n_0 \rangle \langle n_k \V\prod_{i=1}^{k}
\lvec{n_{i-1}} X_{L_i}(N_i) \rvec{n_{i}}
\end{align}
where $X_l(N)$ has been defined in (\ref{PLmatrix}).

In the large $L$ limit (with $q=1-\frac{\lambda}{L}$), (\ref{rhoa}) and (\ref{rhob}) become
\begin{align} \label{largeLrho}
\rho_a &= \frac{\alpha}{\alpha+\gamma} +O(\frac{1}{L}) &
\rho_b &= \frac{\delta}{\beta+\delta}+O(\frac{1}{L})
\intertext{and  (\ref{e}), (\ref{d})}
d&= 1+  O(\frac{1}{L})
& e &= 1 +O(\frac{1}{L})\ \ ,
\label{largeLde}
\end{align}
To compute (\ref{discretwalk}) when $1 \ll L_i \ll L$, let us  evaluate $\lvec{n}
X_l(N)\rvec{n'}$ when  $1\ll l \ll L$. 
When $n$, $n'$ are of order $L$ with $|n-n'|\leq l\ll L$,
we see that for $m$ and $m'$ in $(\min(n,n')-l, \max(n,n')+l)$  all the
non-zero elements of matrix $\lvec{m} \D \rvec{m'}$ and
$\lvec{m'}\E\rvec{m}$ are equivalent to
\begin{equation} 
\lvec{m} \D \rvec{m'} \sim \lvec{m'}\E\rvec{m} \sim
\left(1-e^{-\frac{\lambda n}{L}}\right)\frac{L}{\lambda} \ \ \ .
\end{equation}
The computation of $\lvec{n} X_l(N)\rvec{n'}$ is thus  reduced to an
enumeration problem. This leads to
\begin{align} \label{3.17}
 \lvec{n}
X_l(N)\rvec{n'} &\simeq  \left[\left(1-e^{-\frac{\lambda
n}{L}}\right)\frac{L}{\lambda} \right]^l \sum_{n_+-n_-=n'-n}
\binom{l}{N} \binom{N}{n_+}\binom{l-N}{n_-}
\end{align}
where $\binom{l}{N}$ is the number of words of length $l$ with $N$
matrices $\D$ and $l-N$ matrices $\E$, $n_+$ is the number of matrix
elements of the form $\langle m | \D |m+1 \rangle$ and $n_-$ the
number of matrix elements of the form $\langle m | \E |m-1
\rangle$.\\
Looking at the values of $n_+$ and $n_-$ which dominate (\ref{3.17}) one
obtains
\begin{multline} \label{logFlN}
\log \lvec{n} X_l(N) \rvec{n'} = l \left[ -\log\frac{\lambda}{L}
+\log\left(1-e^{-\lambda y}\right)- \rho \log \rho -(1-\rho) \log
(1-\rho)
\vphantom{\left(1-y'\right) \log \left({1
-y'}\right) }
 \right.\\
-\left(1-\rho+y'\right) \log
\left(1-\rho+y'\right)
\left.
-\left(\rho-y'\right) \log \left(\rho -y'\right)
\vphantom{ log\left(1-e^{-(\lambda y
)}\right)}
+o(1) \right]
\end{multline}
where $y$, $y'$ and $\rho$ are defined by
\begin{align}
y&=\frac{n}{L}& y'&=\frac{n'-n}{l}& \text{and} && \rho=\frac{N}{l}\ \ .
\end{align}

As $\frac{L_i}{L}\to 0$ while $L_i \to \infty$, one can associate to each
walk $w$ a continuous function $y(x)$
\begin{align} \label{defy}
y\left(x_i\right)&=\frac{n_{i}(w)}{L} 
\end{align}
with $x_i=\sum_{j=1}^{i} \frac{L_j}{L}$.
For each walk
$y(x)$ and each density profile $\rho(x)$ such that
\begin{align} \label{cond_walk}
y(x)&> 0 & \text{ and } && \rho(x)-1&\leq y'(x) \leq \rho(x)
\end{align}
(these restrictions on the path $y$ come from $n_i(w)>0$ (see
(\ref{con_disc_walk})) and the condition for $\lvec{n} X_l(N)\rvec{n'}\neq
0$ (see (\ref{logFlN}))), let us define $\Gm{
\{y(x)\},\{\rho(x)\}}$ by
\begin{multline} \label{contweight}
\Gm{ \{y(x)\},\{\rho(x)\}}= y(0) \log\frac{\rho_a}{1-\rho_a}+y(1)
        \log \frac{1-\rho_b}{\rho_b} +\\
         \int_{0}^{1}dx \left[-\log \frac{1-e^{-\lambda y}}{\lambda}+\rho
\log \rho +(1-\rho)\log{(1-\rho)} 
\vphantom{-\log \frac{1-e^{-\lambda y}}{\lambda}+\rho
          \log \rho
        +(1-\rho)\log{(1-\rho)}+(1-\rho+y')
        \log{(1-\rho+y')}+(\rho-y')\log{(\rho-y')}
         }
\right.\\
\left.
\vphantom{-\log \frac{1-e^{-\lambda y}}{\lambda}+\rho
          \log \rho
        +(1-\rho)\log{(1-\rho)}+(1-\rho+y')
        \log{(1-\rho+y')}+(\rho-y')\log{(\rho-y')}
         }
  +(1-\rho+y')
        \log{(1-\rho+y')}+(\rho-y')\log{(\rho-y')} \right]\ \ .
\end{multline} 
Then expression (\ref{discretwalk}) leads to (after replacing the sum
over $w$
by a sup over $w$)
\begin{align} 
\log q_{L_1,L_2,\cdots
L_k}\big(N_1,N_2, \cdots,N_k\big) &\simeq \log P_L\left(\{\rho(x)\}\right)  
\\&\simeq L \sup_{y}
\big[\KWASEP-\Gm{\{y(x)\},\{\rho(x)\}} \big] \label{PLGMK}
\end{align}
with $\KWASEP$ given by
\begin{align} \label{KZL}
\KWASEP &= \lim_{L \rightarrow \infty} \left( \log L -\frac{\log
Z_L(1-\frac{\lambda}{L})}{L}\right) \ \ .
\end{align}
which is the expression (\ref{FWASEPwithy}).

In order to compute $\KWASEP$, we could use a similar method to estimate
$Z_L(q)$ in (\ref{KZL}); we will rather use the property that the most
likely profile $\rhobar$ must verify $\F(\{\rhobar(x)\}; \rho_a,\rho_b)=0$, so we delay the
computation of $\KWASEP$ until the end of section 3.5. Of course both
methods give the
same result.

\subsection{Derivation of  (\ref{FWASEPwithf})  \label{der_FWASEPwithf}}

 As the functions $ -\log(1-e^{-\lambda y(x)})$, $\left(\rho(x)-y'(x)\right) \log
\left(\rho(x)-y'(x)\right)$ and
$ \left(1-\rho(x)+y'(x)\right) \log
\left(1-\rho(x)+y'(x)\right)$  are convex functions of $y$ for every value of $x$ , the function
$\Gm{\{y(x)\},\{\rho(x)\}}$ in (\ref{contweight}) is a sum of convex functions, and is thus
convex.\\
Therefore  there is a unique walk $\ymax(x)$  which minimizes
 $\Gm{\{y(x)\},\{\rho(x)\}}$, so
\begin{align} \label{FG}
\F \left( \{\rho(x)\}; \rho_a,\rho_b \right) &=\Gm{\{\ymax(x)\},\{\rho\}}-\KWASEP
\end{align}
This minimum is not reached at the boundary of (\ref{cond_walk}). Thus
$\ymax(x)$ is the unique stationary point, solution of
$\left. \frac{\partial \Gm{\{y\}, \{\rho\}}}{\partial y(x)}
\right|_{y=\ymax}=0$ and it satisfies:
\begin{align} \label{ymax}
\frac{\rho'(x)-\ymax''(x)}{(1-\rho(x)+\ymax'(x))(\rho(x)-\ymax'(x))}-\lambda \frac{e^{-\lambda
\ymax(x)}}{1-e^{-\lambda \ymax(x)}} &= 0 
\end{align}
with the boundary conditions
\begin{align}
1-\rho(0)+\ymax'(0)&= \rho_a&1-\rho(1)+\ymax'(1)&=\rho_b
\end{align}
in addition to the general  condition (\ref{cond_walk}).

To obtain an expression similar to (\ref{simple_expression}),
we  rewrite (\ref{ymax}) using the function $F(x)$  defined by
\begin{align}
F(x)&=1-\rho(x)+\ymax'(x)  \ \ .\label{Fdef}
\end{align}
This lead to 
\begin{align}
\frac{F'(x)}{F(x)(1-F(x))}&=-\lambda \frac{e^{-\lambda
\ymax(x)}}{1-e^{-\lambda \ymax(x)}}
\label{eqFy}
\end{align}
with the boundary conditions
\begin{align}
F(0) &= \rho_a  &
F(1) &= \rho_b \label{CFF}
\end{align}
The conditions on the walk (\ref{cond_walk}) imply
\begin{equation}
 0\leq F(x) \leq 1 \label{cond_F}
\end{equation}
and as  $y>0$ (see (\ref{cond_walk}) and (\ref{eqFy})) 
\begin{equation} 
F'(x)< 0 \label{Fdecr}
\end{equation}
By eliminating $y$ between (\ref{Fdef}) and  (\ref{eqFy})  we get 
(\ref{f_WASEP}).

\bigskip

The expression of the large deviation functional can be rewritten in
terms
of
$F$ instead of $\ymax$ in (\ref{FWASEPwithy}). 
From (\ref{eqFy}), we see that
\begin{align}
\log\left(1-e^{-\lambda \ymax} \right)&=\log \frac{\lambda F
(1-F)}{\lambda F(1-F)-F'}
\end{align}
If we integrate by part the term $\int_0^1 dx y'(x) \log
\frac{1-\rho(x)+y'(x)}{\rho(x)-y'(x)}$ in (\ref{FWASEPwithy}), we get
\begin{align}
\int_0^1 dx\, y' \log
\frac{1-\rho+y'}{\rho-y'}&=\left[y\log
\frac{1-\rho+y'}{\rho-y'}\right]_0^1 \notag \\
&\ \ \ \ \ +\int_0^1 dx 
\frac{y(\rho'-y'')}{(1-\rho+y')(\rho-y')}\\
&=y(1) \log \frac{\rho_b}{1-\rho_b}-y(0)
\log\frac{\rho_a}{1-\rho_a} \notag \\
&\ \ \ \ \ +\int_0^1 \frac{F'}{\lambda F(1-F)}\log
\frac{-F'}{\lambda F(1-F)-F'}
\end{align}
and thus  (\ref{FWASEPwithy}) leads to 
(\ref{FWASEPwithf}).

Let us now justify (\ref{Ffmax}).
If we define 
\begin{multline} \label{Gbisdef}
\Gbis{\{\rho\}, \{\phi\}}=-\KWASEP+\int_0^1 \left\{ \rho(x) \log
\rho(x)+\big(1-\rho(x)\big)\log\big(1-\rho(x)\big) 
\vphantom{\left(1-\frac{\phi'(x)}{\lambda}
\right) \log (\lambda-\phi'(x))} 
\right. \\ +\big(1-\rho(x)\big)\phi(x)-\log\left(1+e^{\phi(x)}\right)
\\ \left. +\frac{\phi'(x)}{\lambda}\log\big(-\phi'(x)\big)+\left(1-\frac{\phi'(x)}{\lambda}
\right) \log \big(\lambda-\phi'(x)\big)\right\} \ \ ,
\end{multline}
expression (\ref{FWASEPwithf}) implies that when $F$ is solution of
(\ref{f_WASEP}) with condition (\ref{f_WASEPBC}) then
\begin{align} \label{FGbis}
\F\big(\{\rho\}; \rho_a,\rho_b \big)&=\Gbis{\{\rho\},
\left\{\log\left(\frac{F}{1-F}\right)\right\}}
\end{align}
As $-\log(1+e^{\phi(x)})$ and $\frac{\phi'(x)}{\lambda}\log\big(-\phi'(x)\big)+\left(1-\frac{\phi'(x)}{\lambda}
\right) \log \big(\lambda-\phi'(x)\big)$ are concave function of $\phi$
for $\phi'<0$,
$\Gbis{\{\rho\}, \{\phi \}}$ is a concave
function of $\phi$. So (\ref{f_WASEP}) is
equivalent to the condition for $F$ to maximize
$\Gbis{\{\rho\},\left\{\log\left(\frac{F}{1-F}\right)\right\}}$ under the constraint
(\ref{f_WASEPBC}) and $F'<0$, i.e. 
\begin{align}
\frac{\partial \Gbis{\{\rho\},
\left\{\log\left(\frac{F}{1-F}\right)\right\}}}{\partial F}&=0 \ \ .
\end{align} So (\ref{FGbis}) can be written as
\begin{align} \label{FGbismax}
\F\big(\{\rho(x)\};\rho_a, \rho_b\big)&= \sup_F \left[\Gbis{\{\rho\},
\left\{\log\left(\frac{F}{1-F}\right)\right\}}\right]
\end{align}
where the $\sup$ is taken over decreasing functions $F$ with the condition
(\ref{f_WASEPBC}), leading to (\ref{Ffmax}).

\subsection{The most likely profile \label{mlp}}

From (\ref{Ffmax}), we get 
\begin{align}
\frac{\partial \F(\{\rho(x)\};\rho_a, \rho_b)}{\partial \rho(x)}
&=\log\frac{\rho (1-F)}{(1-\rho)F }
\end{align}
so that for the  most likely profile $\rhobar(x)$
\begin{align} \label{frhobar}
F(x)&=\rhobar(x)\ \ .
\end{align}
Equation (\ref{f_WASEP}) thus becomes
\begin{align} \label{eqrhobar2}
\rhobar''=\lambda \rhobar' (1-2 \rhobar)
\end{align}
with boundary conditions
\begin{align} \label{rhobarbc}
\rhobar(0)&=\rho_a& \text{and}&&\rhobar(1)&=\rho_b
\end{align}
Integrating (\ref{eqrhobar2}) once, we get constant parameter $J$ 
\begin{align} \label{eqrhobar1}
\rhobar'=\lambda \rhobar (1-\rhobar)-J
\end{align}
The boundary conditions (\ref{rhobarbc}) determine $J$ as given by
equation (\ref{current}) and $\rhobar$ by (\ref{solrhobar}).

One can show that (\ref{FWASEPwithf}) implies that
\begin{align}
<\tau_i \tau_{i+1}> &= <\tau_i> <\tau_{i+1}>+O\left(\frac{1}{L}\right)
\end{align}
(see for example \cite{DLS2} for the case $\lambda = 0$).\\
In particular $<\tau_i (1-\tau_{i+1})> \simeq <\tau_i> <1-\tau_{i+1}>$
and (\ref{j_corr_finc}) for
$q=1-\frac{\lambda}{L}$ leads to
\begin{align} \label{jchpmoy}
j&= \frac{\lambda}{L} \rhobar
(1-\rhobar)-\frac{\rhobar'}{L}+o\left(\frac{1}{L}\right)\ \ .
\end{align}
Comparing (\ref{jchpmoy}) to (\ref{eqrhobar1}) leads to relation (\ref{Jdef})
between $J$ and the current $j$.
 
Depending on the reservoir densities $\rho_a$ and $\rho_b$, one gets
various expression for the most likely profile $\rhobar(x)$
\begin{itemize}
\item when
$\rho_a-\rho_b > \lambda
\left(\rho_a-\frac{1}{2}\right)\left(\rho_b-\frac{1}{2}\right)$
then 
$J>\frac{\lambda}{4}$: 
\begin{align} \label{tanpro}
\rhobar(x)&=\frac{1}{2}-\sqrt{\frac{J-\lambda/4}{\lambda}} \tan
\left[\sqrt{\lambda(
J-\lambda/4)}(x-x_0)\right]
\end{align}
\item and  when
$\rho_a-\rho_b<\lambda\left(\rho_a-\frac{1}{2}\right)\left(\rho_b-\frac{1}{2}\right)$ then
$J<\frac{\lambda}{4}$ and
\begin{align} \label{cothpro}
\rhobar(x)&=\frac{1}{2}+\sqrt{\frac{\lambda/4-J}{\lambda}} \coth \left[
\sqrt{\lambda (
\lambda/4-J)}(x-x_0)\right]
\end{align}
\end{itemize}
where $x_0$ and $J$ are chosen to satisfy (\ref{rhobarbc}).

Starting with the expression of the most likely profile, one can get an equation
for $\KWASEP$ by writing that
\begin{align} \label{Frhobar}
\F(\{\rhobar\};\rho_a,\rho_b)=0
\end{align} 
Introducing equations (\ref{frhobar}) and (\ref{eqrhobar1}) in expression
(\ref{FWASEPwithf}) and solving (\ref{Frhobar}) we get expression (\ref{KWASEPJ}) for $\KWASEP$.

\section{Limiting cases}

In this section we show how previously known expressions
(\ref{simple_expression})-(\ref{Kdef1}) can be recovered as limiting
cases of the results (\ref{FWASEPwithy}-\ref{solrhobar}) of the present
paper.

\subsection{The SSEP limit}

Let us first consider the small $\lambda$ limit.
Expression (\ref{current}) for the current can be expanded in powers of
$\lambda$. 

\begin{align}
j&=\frac{\rho_a-\rho_b}{L} +\frac{\lambda}{L} \left(\frac{\rho_a+\rho_b}{2}-\frac{\rho_a^2+\rho_a
\rho_b+\rho_b^2}{3}\right) +O(\lambda^2)
\end{align}
in agreement (when $\lambda \to 0$) with $j=\frac{\rho_a-\rho_b}{L}$ of
\cite{HS}.\\
The large deviation functional (\ref{FWASEPwithf})  
becomes for small
$\lambda$ 
\begin{multline}
\F=\int_0^1 dx \left\{\rho  \log
\left(\frac{\rho}{F}\right)+(1-\rho)\log
\left(\frac{1-\rho}{1-F}\right) \right. \\\left. \vphantom{\rho 
\log
\left(\frac{\rho}{F}\right)+(1-\rho)\log
\left(\frac{1-\rho}{1-F}\right)}+\log(-F')-\frac{\lambda F
(1-F)}{2 F'}\right\}+1-\KWASEP+O(\lambda^2)\ \ .
\end{multline}
The leading order in $\lambda$ agrees with the SSEP expression (\ref{simple_expression}) as the
constant $\KWASEP$  given by (\ref{KWASEPJ}) becomes
\begin{align} 
\KWASEP&= \log(\rho_a-\rho_b)+1+O(\lambda)
\end{align}
and the equation (\ref{rho(t)}) for $F$ is the limiting case of 
(\ref{f_WASEP}) when $\lambda\to 0$.
Furthermore one can check that  the most likely
profile (\ref{solrhobar}) becomes linear as  $J=Lj \to \rho_a-\rho_b$  in the small $\lambda$ limit.
Thus  the results known for SSEP are recovered from our general
$\lambda$ case in the limit $\lambda \to 0$
.
\subsection{The ASEP limit for $\rho_a>\rho_b$}

Let us now see how
the strongly asymmetric case  (\ref{result}) can be recovered from
(\ref{FWASEPwithf})  for large $\lambda$.
We see that for large $\lambda$, the solution $J$ of (\ref{current}) is
$J\simeq \max\limits_{\rho_b \leq \rho \leq \rho_a} \lambda \rho
(1-\rho)$, so that 
\begin{align} \label{lllcurent}
j&\sim \sup\limits_{\rho_b \leq \rho \leq \rho_a}\frac{ \lambda}{L} \rho (1-\rho) &&
\text{when} & \rho_a>\rho_b
\end{align}
in agreement with \cite{PS, PSbis}, and  that $\KWASEP \simeq \log J$ in
(\ref{KWASEPJ}) so that
(\ref{Ffmax}) reduces to (\ref{result}).

An interesting aspect of this large $\lambda$ limit is to
see how the solution $F$ of (\ref{f_WASEP}) becomes for large
$\lambda$ the function $F$ constructed from $1-\rho(x)$ by the Maxwell
construction explained after equation (\ref{Kdef1}).
When $\lambda$ becomes large,  (\ref{f_WASEP}) implies that $F \simeq
1-\rho$ or $F' \simeq 0$. Therefore one expects a succession of domains
where $F(x) \simeq 1-\rho(x)$ and domains where $F(x)$ is constant, with
the remaining contraints that $F$ is monotone and $\rho_b \leq F(x) \leq
\rho_a$

In a domain where  $F(x)\simeq 1-\rho(x)$, one can expand $F$ 
in powers of $\frac{1}{\lambda}$
\begin{align}
F&=1-\rho+\frac{1}{\lambda}\left[\rho'\frac{1-2\rho}{\rho(1-\rho)}-\frac{\rho''}{\rho'}\right]+O\left(\frac{1}{\lambda^2}\right)
\ \ \ .
\end{align}
The condition $F'<0$ (\ref{Fdecr})   implies that in such domains  $\rho'>0$

In  a domain $t<x<u$ where $F(x)\simeq C$ is almost constant, neglecting
${F'}^2$ in
 (\ref{f_WASEP}) (as ${F'}^2 \ll \lambda F'$) gives
\begin{align} \label{F'const}
F'(x)&= B \exp{\left[-\lambda \int^x_t{(C-1+\rho(x'))dx'}
\right]}
\end{align}
where $B$ is constant over the whole domain $t<x<u$.

The next question is to understand for large $\lambda$ the transitions
between these different domains. 

A first possibility is that $F(x)$ (which is monotone) becomes
discontinuous at such transition point. This is the case for example when
$1-\rho(x)$ is decreasing and discontinuous, implying that $F(x)$ has a
variation of the order of $1$ (the discontinuity of $\rho$) over a range
in $x$ of order of $\frac{1}{\lambda}$. This case can be analysed without
difficulty, but we won't discuss it here.

\begin{figure}[ht!]
\begin{center}% \label{figrhodisc}
\includegraphics{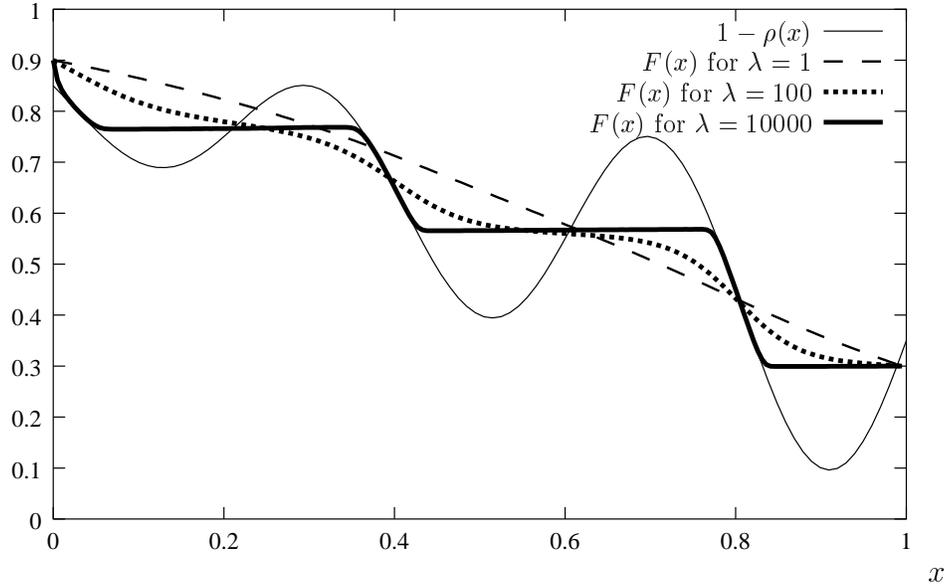}
\end{center}
\caption{$F(x)$ solution of (\ref{f_WASEP}) versus $x$ for a given profile $\rho(x)$ and
reservoir densities $\rho_a=0.9$, $\rho_b=0.3$ when $\lambda$ takes the value $1$,
$100$ and $10000$. The full-curve is $1-\rho(x)$. The curves are obtained
by minimizing (\ref{FWASEPwithy}) numerically  for $y(x)$ (discretized over
$200$ points).  $F(x)$ is then calculated from $y(x)$ by relation
(\ref{Fdef}). For large $\lambda$, we see a succession of domains where
$F=1-\rho(x)$ and where $F$ is constant satisfying the Maxwell
construction.
\label{figrhodisc}
}
\end{figure}

The other possibility is that for large $\lambda$, $F$ remains continuous
but $F'$ becomes discontinuous. This is what happens for example on figure
\ref{figrhodisc}: for large $\lambda$ there is a succession of domains
where $F=1-\rho(x)$ and where $F$ is constant.

Let us consider a domain where $F(x)=1-\rho(t)=1-\rho(u)$ for $t<x<u$ in
the large $\lambda$ limit, surrounded by two domains where
$F(x)=1-\rho(x)$.
At the transition point $t$, the solution of (\ref{f_WASEP}) takes a
scaling form 
\begin{align}
F(x)-1+\rho(x)&=\sqrt{\frac{\rho'(t)}{\lambda}} G\big((x-t) \sqrt{\rho
'(t)
\lambda}\big)
\end{align}
 where $G$ is 
solution of
\begin{align}
G''&=G (1-G') \label{point_tournant}
\end{align}
For $F$ to match the solution $1-\rho(x)+O(\frac{1}{\lambda})$ for $x=t-0$
and $1-\rho(t)+o(1)$ for $x=t+0$, one needs that 
\begin{align}
G(z)&\rightarrow 0 &&\text{when} &z\rightarrow -\infty \label{CLpt}\\
G(z)&=z+o(1) &&\text{when} &z\rightarrow \infty \label{CRpt}
\end{align}
It can be shown  that then $G$ is solution of
\begin{align}
G'&=1+W\left(A e^{-\frac{G^2}{2}}\right)
\end{align}
where $W$ is the product logarithm function (also called the Lambert
function, see \cite{Knuth}) defined here as the largest real solution of
\begin{align}
W(x)e^{W(x)}=x
\end{align}
Condition (\ref{CLpt}) determines  $A=-e^{-1}$ (as
$W\left(-e^{-1}\right)=-1$).
Using that for $x$ small $W(x)\simeq x$, the limit of $G$  for large
$z$ can be
computed
\begin{align}
G(z)-z &\sim \frac{1}{e} \int_z^\infty e^{-\frac{{z'}^2}{2}}\, dz'
\end{align}
It gives  for the asymptotic regime (when $\frac{1}{\sqrt{\lambda}}\ll x-t \ll 1$) of $F'$  
\begin{align} \label{Gasl}
F'(x) \sim -\frac{\rho'(t)}{e}e^{-\frac{\rho'(t) \lambda}{2}(x-t)^2}
\end{align}
At the other boundary $u$ of the domain, $F$ has a similar scaling form.
\begin{align}
F(x)-1+\rho(x)&=-\sqrt{\frac{\rho'(u)}{\lambda}} G\big(-(x-u) \sqrt{\rho
'(u)
\lambda}\big)
\end{align}
with the same $G$ solution of (\ref{point_tournant}).
This gives  for the asymptotic regime $\frac{1}{\sqrt{\lambda}}\ll-
(x-u)
\ll 1$ 
\begin{align} \label{Gasr} 
F'(x) \simeq -\frac{\rho'(u)}{e}e^{-\frac{\rho'(u) \lambda}{2}(x-u)^2}
\end{align}
For the asymptotics of (\ref{F'const}) to match with (\ref{Gasl}) as
$x\to t$
\begin{align}
F'(x) \sim Be^{-\frac{\rho'(t) \lambda}{2}(x-t)^2}
\end{align}
and with (\ref{Gasr}) when $x\to u$
\begin{align} \label{4.20}
F'(x) \sim Be^{\left[-\lambda \int^u_t{(C-1+\rho(x'))dx'}
\right]-\frac{\rho'(u) \lambda}{2}(x-u)^2}
\end{align}
 one
needs  that to the leading order in $\lambda$ 
\begin{align} \label{4.22}
\int^u_t(C-1+\rho(x'))dx' &=0
\end{align}
which is the Maxwell construction  (\ref{maxwell}).
We see that the constant $C$, the value of $F(x)$ in a domain where $F$
is constant, is determined by an expression (\ref{4.22}) which is obtained
from two matching conditions at the boundaries of the domain. This is
very similar to the Bohr Sommerfeld rule  which determines the energy
levels in the WKB method \cite{WKB,MathMet}. So the Maxwell construction
here has a mathematical origin similar to  the Bohr Sommerfeld rule.

\section{Extension of  our results
\label{extension}}
\subsection{Extension of the domain (\ref{val_dom1})}

We are going to show that  our results of
section 2, initially derived in the domain (\ref{val_dom1}) remain valid
for
\begin{align} \label{cond2bis}
\lambda&>0 &    e^{\lambda}\frac{1-\rho_b}{\rho_b} \frac{\rho_a}{1-\rho_a}
&<1  
\end{align}

The representation of the algebra for $\D$, $\E$, $\V$ and $\W$
introduced
in section 3.2 
remains valid for some range of the parameter $\rho_a$ and $\rho_b$ when
$q>1$. Indeed, for large $n$ the non-zero  matrix elements of the kind
 $\lvec{n} \D \rvec{n'}$  and $\lvec{n'} \E \rvec{n}$ behave then  like
$q^{n}$. Furthermore, $\frac{(ed;q)_n}{(q;q)_n} \sim
\left(\frac{ed}{q}\right)^n$ for large $n$. Thus, using
(\ref{d}) and (\ref{e}), we get that for any product $X$ of $L$ matrices $\D$
or $\E$ (and any sum of such product)
$\W n\rangle\langle n| X \V \sim
\left(\frac{\gamma \delta}{\alpha \beta} q^{L-1}\right)^n$
and the condition for $\W X\V$ to be finite is 
\begin{align} \label{limrep2}
\frac{\gamma \delta}{\alpha \beta} q^{L-1}&<1 \ \ .
\end{align}

When $q=1-\frac{\lambda}{L}$ and in the large $L$ limit, this leads to
(using (\ref{largeLrho}) )
\begin{align} \label{cond2}
\lambda&<0 &    e^{-\lambda}\frac{1-\rho_a}{\rho_a}
\frac{\rho_b}{1-\rho_b}
&<1  
\end{align}
So all the content of sections 3.3 to 3.6 remains valid, leading
thus to formulas (\ref{FWASEPwithy})-(\ref{solrhobar}). The only change
is that in (\ref{Ffmax}), the $\sup$ is now over decreasing
functions $F$
such that for every $x$, $\lambda F(x) \big(1-F(x)\big) - F'(x)>0$.

In order to recover the expressions for $\lambda>0$, we use the
left-right symmetry of the system, replacing $\lambda$ by $-\lambda$, $x$
by $1-x$, $\rho_a$ by $\rho_b$, \dots so that condition (\ref{cond2}) becomes (\ref{cond2bis}). 

When condition (\ref{cond2bis}) is fulfilled, $J$ given by
(\ref{current}) is negative, and thus there is a current $j$
going from the reservoir with the highest density $\rho_b$ to the
reservoir with the lowest density $\rho_a$, despite
the external bias $q$.
The most likely profile, solution of (\ref{eqrhobar1}) and
(\ref{rhobarbc}) is now given by
\begin{align} \label{thpro}
\rhobar(x)&=\frac{1}{2}+\sqrt{\frac{\lambda/4-J}{\lambda}} \tanh
\sqrt{\lambda(
\lambda/4-J)}(x-x_0)
\end{align}

\subsection{The detailed balance case}

We show now  that (\ref{FWASEPwithf}) remains also
valid
when the boundary parameters $\alpha$, $\beta$, $\gamma$ and $\delta$
are such that detailed balance is satisfied.
Detailed balance means that the probability of observing a transition
from a microscopic configuration ${\cal C}$ to another ${\cal C}'$ is equal to the probability of
observing the reversed transition  (from ${\cal C}'$ to ${\cal C}$). 

Let $\{\tau_i\}$ be the occupation numbers of a given microscopic
configuration. Detailed balance corresponding to a jump of a particle
between sites $k$ and $k+1$ means that
\begin{align}\label{detbalk}
P\big(\{\tau_1,\ldots,\tau_{k-1},0,1,\tau_{k+2}, \ldots, \tau_L\}\big)&=q^{-1}
P\big(\{\tau_1, \ldots,\tau_{k-1},1,0,\tau_{k+2}, \ldots, \tau_L\}\big)
\end{align}
where $P\big(\{\tau_i\}\big)$ is the steady state probability of
configuration $\{\tau_i\}$.\\
The detailed balance relation at the left boundary is
\begin{align} \label{detbal1}
P\big( \left\{1, \tau_2,\ldots, \tau_L\right\} \big) &=
\frac{\alpha}{\gamma}P\big( \left\{0,
\tau_2, \ldots, \tau_L\right\}\big)
\end{align}
and at the right boundary 
\begin{align} \label{detbalL}
P\big( \left\{ \tau_1,\ldots, \tau_{L-1},1\right\} \big) &=  \frac{\delta}{\beta}P\big(
\left\{
\tau_1, \ldots, \tau_{L-1},0 \right\}\big)
\end{align}
Starting from a configuration with occupation number $\{\tau_i\}$, one can
always use (\ref{detbalk}) and
(\ref{detbal1}) to calculate the weights of all configurations by
removing particles at the left boundary.
\begin{align} \label{recl}
P\big( \{\tau_i\}\big)&= \prod_{i=1}^L
\left(\frac{\alpha}{\gamma q^{i-1} }\right)^{\tau_i} P\big( \left\{
0,0,\ldots,0\right\} \big)
\end{align}
For general values of $\alpha$, $\beta$, $\gamma$ and $\delta$, these
weights do not satisfy (\ref{detbalL}) and are not steady state weights.
If we insist however that (\ref{recl}) satisfies also (\ref{detbalL}), we get
\begin{align} \label{conddetbal}
q^{L-1} \frac{\gamma \delta}{\alpha \beta}&=1
\end{align}
which is the detailed balance condition, and if (\ref{conddetbal}) is
satisfied, we get
\begin{align} \label{probdetbal}
P\left(\{\tau_i\}\right) &= \frac{\prod_{i=1}^L
\left(\frac{\alpha}{q^{i-1} \gamma}\right)^{\tau_i}}{\prod_{i=1}^L
\left(1+\frac{\alpha}{q^{i-1}\gamma}\right)}
\end{align}
The average density at site $i$ is thus given by
\begin{align} \label{detbalrho}
 \rhobar\left(\frac{i}{L}\right)&=\frac{1}{1+q^{i-1}\frac{\gamma}{\alpha}}
\end{align}

In the weak asymmetry regime (\ref{WASEPdef}), expression
(\ref{conddetbal}) and (\ref{detbalrho}) become
\begin{align} \label{detbalweakcond}
e^{\lambda } \frac{1-\rho_b}{\rho_b}\frac{\rho_a}{1-\rho_a}&= 1
\end{align}
and 
\begin{align} \label{rhodbw}
\rhobar(x)&= \frac{\rho_a}{\rho_a+(1-\rho_a) e^{-\lambda x}}
\end{align}
whereas (\ref{probdetbal}) leads to the following expression for the
large deviation functional
\begin{align} \label{Fdetbalrho}
\F(\{\rho(x)\};\rho_a,\rho_b)&=\int_0^1 \left\{\rho \log
\frac{\rho(x)}{\rhobar(x)}+(1-\rho(x))\log
\left(\frac{1-\rho(x)}{1-\rhobar(x)}\right) \right\}dx
\ \ \ \ .
\end{align}
Note that (\ref{detbalweakcond}) corresponds to the boundary of the range
of parameters (\ref{cond2bis}) where we have shown (\ref{FWASEPwithf}) to be
valid.
Although our derivation of (\ref{FWASEPwithf}) was not a priori valid
when detailed balance
(\ref{detbalweakcond}) is satisfied (see (\ref{cond2bis})), we are going to
see now that (\ref{rhodbw}) and (\ref{Fdetbalrho}) can nevertheless be
recovered.\\
When detailed balance (\ref{detbalweakcond}) holds, one can check that 
\begin{align} \label{detbalrhobarf}
F(x)=\rhobar(x)&=\frac{\rho_a}{\rho_a+(1-\rho_a)e^{-\lambda x}}
\end{align}
is solution of (\ref{f_WASEP}) for arbitrary $\rho(x)$. So
(\ref{detbalweakcond}) implies that $F(x)$ does not depend on $\rho(x)$.
As $F(x)$ given by (\ref{detbalrhobarf}) satisfies 
\begin{align}
F'&=\lambda F(1-F) \ \ ,
\end{align}
one can see that (\ref{FWASEPwithf}) reduces to (\ref{Fdetbalrho}).\\
On can also check that when detailed balance is satisfied the current
(\ref{current},\ref{Jdef}) vanishes and that $J=0$ in (\ref{current}) is
equivalent to (\ref{detbalweakcond}).

\section{Conclusion}

In the present work, we have obtained the expressions (\ref{FWASEPwithy},
\ref{FWASEPwithf}, \ref{Ffmax}) for the large deviation functional of the
one dimensional simple exclusion process in the weak asymetry regime
(\ref{WASEPdef}). Our analysis of the limiting cases ($\lambda \to 0$ and
$\lambda \to \infty$) has shown that these new expressions are consistent
with previously known expressions for the SSEP and the ASEP.

For technical reasons, our derivation is limited to some ranges of
parameter (\ref{val_dom1}) , (\ref{cond2bis}) or (\ref{detbalweakcond}). It
would of course be useful to know what happens in the other ranges of
parameters, if  our results remain valid or not and how the ASEP result
(\ref{result3}) can be recovered.

The derivation of our results, based on the matrix representation of the
steady state, uses strongly that the steady state weights of the
configurations can be written as sums over paths of the weights of these
paths. We used a similar idea recently to study the density fluctuations
in the TASEP \cite{DEL}. These paths have so far a purely mathematical
origin and it would be of course interesting to give them a physical
interpretation.

Another interesting question would be to see whether the results of the
present paper could be understood using the macroscopic fluctuation
theory \cite{BDGJL, BDGJL2}. 

\appendix

\section{Appendix: Definition of the densities $\rho_a$ (\ref{rhoa}) and
$\rho_b$ (\ref{rhob}) of the reservoir}

When the boundary parameters $\alpha$, $\beta$, $\gamma$ and $\delta$
satisfy a certain relation ((\ref{6.3bis}) below), the steady state is a
Bernoulli measure at density $\rho$ and one can consider that the two
reservoirs are at this same density $\rho$.

When the steady state  is a Bernoulli measure
at density $\rho_a$, the steady state current (\ref{j_corr_finc}) in the bulk is given by
\begin{align} \label{jbulk}
j&=(1-q) \rho_a (1-\rho_a)
\end{align}
and at the left boundary, by
\begin{align} \label{jleft}
j&=\alpha (1-\rho_a)-\gamma \rho_a
\end{align}
The conservation of  particles implies that (\ref{jbulk}) and
(\ref{jleft}) should coincide and this gives condition (\ref{eq_rhoa})
for $\rho_a$. If one repeats the same argument at the right boundary, one
gets that $\rho_b$ should satisfy 
\begin{align}
(1-q) \rho_b (1-\rho_b)&=\beta \rho_b-\delta (1-\rho_b)
\end{align}
leading to equation (\ref{eq_rhob}). The solutions of
(\ref{eq_rhoa}) and (\ref{eq_rhob}) (satisfying $0 \leq \rho_a \leq 1$,
$0 \leq \rho_b \leq 1$ ) are given in (\ref{rhoa}) and (\ref{rhob}).
 One recovers in particular  
$\rho_a=\min(\frac{\alpha}{1-q}, 1)$,
$\rho_b=\max(1-\frac{\beta}{1-q},0)$ when $\gamma=\delta=0$ as in
\cite{DLSasep}, and  $\rho_a=\frac{\alpha}{\alpha+\gamma}$, 
$\rho_b=\frac{\delta}{\beta+\delta}$ when $q \to 1$ as in \cite{DLS2}.
Comparing (\ref{eq_rhoa}) and (\ref{eq_rhob}), one can check that for the two
reservoirs to be at the same density $\rho$ (i.e. for
$\rho_a=\rho_b=\rho$), the boundary parameters should satisfy 
\begin{align} \label{6.3bis}
(\alpha+\delta) (\beta +\gamma) (1-q) &= (\alpha \beta-\gamma \delta)
(\alpha +\beta+\gamma+\delta) \ \ .
\end{align}

Let us now verify that when $\rho_a=\rho_b=\rho$, the steady state measure is
indeed a Bernouilli measure at density $\rho$. Consider a configuration $\{\tau_i\}$ with the steady state
probability $P$. The probability of leaving this configuration during the
time interval $dt$ is
\begin{align} \label{quit}
\big[ (1+\gamma) \tau_1 +\alpha (1-\tau_1) +n_c (1+q) +\delta (1-\tau_L)
+(\beta +q) \tau_L \big] P \, dt
\end{align}
where $n_c$ is the number of clusters of particles in $\{\tau_i\}$ which do not touch
a  boundary.\\
If the steady state is Bernoulli at density $\rho$, the probability of
entering the configuration $\{\tau_i\}$ is
\begin{multline} \label{enter}
\big[ \tau_1  \left( q+\alpha \frac{1-\rho}{\rho} \right) +\gamma
\frac{\rho}{1-\rho}(1-\tau_1) + n_c (1+q)\\+\tau_L \left(1+\delta
\frac{1-\rho}{\rho}\right) + \beta \frac{\rho}{1-\rho} (1-\tau_L)\big] P \, dt
\end{multline}
For (\ref{quit}) and (\ref{enter}) to be equal for any $\{\tau_i\}$, one needs that
\begin{align}
1+\gamma-\alpha &= q+\alpha \frac{1-\rho}{\rho} -\gamma
\frac{\rho}{1-\rho} \label{rhoeq1}\\
-\delta+\beta+q &= 1+\delta \frac{1-\rho}{\rho} -\beta
\frac{\rho}{1-\rho} \label{rhoeq2}\\
\alpha +\delta &= \gamma \frac{\rho}{1-\rho} +\beta \frac{\rho}{1-\rho}
\label{rhoeq3}
\end{align}
Comparing (\ref{rhoeq1}) and (\ref{rhoeq2}) to (\ref{eq_rhoa}) and
(\ref{eq_rhob}), one sees that
\begin{equation}
\rho=\rho_a=\rho_b
\end{equation}
whereas (\ref{rhoeq3}) is equivalent to the difference of (\ref{eq_rhoa})
and
(\ref{eq_rhob}) so is automatically satisfied.

\section*{Acknowledgments}

 We thank  V.~Hakim, J.~L.~Lebowitz and E.~R.~Speer for useful and
encouraging discussions.

\newpage

\listoffigures

\end{document}